\pdfoutput=1
\documentclass[10pt,twocolumn]{article}

\usepackage{lmodern}
\usepackage[T1]{fontenc}
\usepackage[a4paper, top=2.0cm, bottom=2.4cm, left=1.5cm, right=1.5cm, columnsep=0.65cm]{geometry}
\usepackage{amsmath}
\usepackage{booktabs}
\usepackage{graphicx}
\usepackage{microtype}
\usepackage{titlesec}
\usepackage{caption}
\usepackage{xcolor}
\usepackage{url}
\usepackage{fancyhdr}
\usepackage{balance}
\usepackage{hyperref}

\graphicspath{{figures/}}

\hypersetup{
  colorlinks=true,
  linkcolor=blue!60!black,
  citecolor=blue!60!black,
  urlcolor=blue!60!black,
  pdftitle={Tracing model-generated DNA with position-independent watermarking},
  pdfauthor={Kimon Antonios Provatas; Christos Galanopoulos; Ilias Georgakopoulos-Soares},
  pdfsubject={Position-independent detection of the SynthID tournament watermark in two genomic language models},
  pdfkeywords={genomic language models; watermarking; SynthID; provenance; position-independent detection}
}
\titleformat{\section}{\normalfont\large\bfseries}{\thesection}{1em}{}
\titleformat{\subsection}{\normalfont\normalsize\bfseries}{\thesubsection}{1em}{}

\begin{document}

\twocolumn[{%
\begin{center}
{\LARGE\bfseries Tracing model-generated DNA\\[3pt]
with position-independent watermarking\par}
\vspace{1.0em}
{\large
Kimon Antonios Provatas\textsuperscript{1,*}\hspace{0.9em}
Christos Galanopoulos\textsuperscript{1,*}\\[4pt]
Ilias Georgakopoulos-Soares\textsuperscript{1,\dag}\par}
\vspace{0.6em}
{\footnotesize
\textsuperscript{1}The University of Texas at Austin, Austin, Texas, USA\\[2pt]
\textsuperscript{*}These authors contributed equally.\quad
\textsuperscript{\dag}Correspondence: \href{mailto:ilias@austin.utexas.edu}{ilias@austin.utexas.edu}\par}
\vspace{1.4em}
\begin{minipage}{0.86\textwidth}
\small
\textbf{Abstract.}
Genomic language models can write synthetic DNA that carries no record of its origin. A
generation-time watermark could provide a provenance signal, if a verifier can detect it using only
the DNA sequence, a key, and published detector settings, without knowing where the generated
region starts, which strand it lies on, or how it was divided into six-base tokens. We embed the SynthID tournament watermark in two genomic language models,
Carbon and GENERator-v2. The verifier searches both strands, every start position, and four window
lengths, and corrects its decision for the whole search. In a detection cohort of 1{,}544 held-out
prompts, it found all 3{,}088 marked sequences per model, unedited and after one substituted, inserted,
or deleted base. For ordinary sequences, the one-sided 95\% upper confidence bound on the
false-positive rate was at most 0.850\%; one bound for the wrong-key control reached 1.015\%
(Carbon, after a deletion). In a development cohort, neither model showed a detectable change in
likelihood or in predefined sequence measures. Under random edits placed without reference to the
detector, detection stayed complete or nearly complete up to a 2\% per-base edit rate. These
results show statistical detection, not biological function, secret-key security, or robustness to
an editor who sees the detector.
\end{minipage} \vspace{1.2em} \end{center}
}]

\thispagestyle{fancy}

\section{Introduction}

Genomic language models now generate coding sequences, regulatory elements, genome-scale
constructs, and functionally constrained DNA designs
\cite{nguyen2024evo,wu2025generator,king2026bacteriophages}. A DNA sequence does not record
whether it came from an organism, a design process, or a generative model, and that provenance is
lost once the sequence leaves the system that produced it. One biosecurity proposal for designed
proteins is to store synthetic gene sequences and synthesis data in repositories that are queried
only in emergencies \cite{baker2024protein}. A provenance signal carried by the sequence itself
would complement such records, because it cannot be separated from the sequence.

Watermarking embeds a statistical signal during generation, and a verifier holding the key can
later test for it without the model, the prompt, or the generation logs
\cite{kirchenbauer2023watermark}. In DNA, the verifier must first find where to look. Many genomic
models write DNA as non-overlapping six-base tokens, but the emitted sequence does not show where
one token ends and the next begins. We call the offset of that division the token phase. There are
six phases, and the term is unrelated to alignment against a reference genome. Reverse
complementation changes the tokens and their order, so both strands must be considered. When the generated region sits inside a longer sequence, its start and end are also
unknown. The verifier has to search all of these possibilities and still control the probability
of a false positive for the sequence as a whole.

We integrate the SynthID tournament watermark \cite{dathathri2024scalable} into two genomic
language models, Carbon-500M \cite{carbon2026,carbon500m} and GENERator-v2 1.2B
\cite{generatorv22026,generatorv2checkpoint}. Both write six-base tokens and can sample directly
from the same vocabulary. They are not independent, because the Carbon training corpus is built
mainly on GENERator-v2's pre-training data \cite{carbon2026}. SynthID preserves the model's output
distribution in expectation over fresh keyed functions, and at a fixed key and context it samples
from a deliberately reweighted distribution. We ask whether the sampler follows that distribution,
whether watermarking changes likelihood or simple sequence statistics, and whether a verifier can
find the mark without knowing the start, strand, or token phase.

The study uses two cohorts of prompts with separate roles. A development cohort of 256 prompts
serves the sampler check, the quality comparison, and the choice of how to score a window. A
detection cohort of 1{,}608 prompts, of which 1{,}544 are held out, serves detection, the
false-positive rate, and the edit experiments.

In both models, the verifier found every marked sequence in the detection cohort, unedited and after
one edited base. For ordinary sequences, every one-sided 95\% upper confidence bound on the
false-positive rate lay below the 1\% target. Neither model showed a detectable change in the
quality measures. Random edits placed without reference to the detector left detection complete or
nearly complete up to a 2\% per-base edit rate. After an insertion or deletion, the windows the
verifier selected are consistent with regions where the token phase is preserved or recovered.

\section{Background}
\label{sec:background}

\subsection{Watermarking designed DNA}

DNA watermarks predate generative models. The first synthetic bacterial genome carried watermark
sequences \cite{gibson2010creation}, and a short signature was written into the genome of an
engineered \textit{Bacillus subtilis} strain \cite{arita2004secret}. DNA-Crypt hid messages in
synonymous codons \cite{heider2007dna}, and messages placed in an open reading frame without
changing the encoded protein persisted during viral replication \cite{liss2012watermarks}. These
designs need a location for the payload and attention to the biological effect of each change. Generation-time
watermarking addresses a complementary setting by embedding the signal through the model's
sampling distribution during generation, so it needs no payload region.

\subsection{Watermarking language-model output}

Language-model watermarks change how tokens are sampled. Some bias a pseudorandom subset of tokens
and detect its excess frequency \cite{kirchenbauer2023watermark}. Others leave the output
distribution unchanged in expectation over fresh keyed randomness
\cite{christ2024undetectable,kuditipudi2024robust,aaronson2022watermark,hu2024unbiased,wu2024dipmark}.
SynthID in its non-distortionary configuration belongs to the second group. We apply it to DNA and
propose no new construction. Robustness depends on what the attacker can do, and prior work
describes evasion, key-recovery, and forgery attacks by stronger and adaptive attackers
\cite{zhang2024sand,diaa2025adaptive,sadasivan2025reliably,jovanovic2024stealing,reynolds2025breaking}.
This study tests none of them; its edits are random and never consult the detector.

Text watermarks built to survive editing treat synchronization as a local problem. Kuditipudi
et al.\ align a candidate text to a key sequence under an edit distance
\cite{kuditipudi2024robust}, and Kirchenbauer et al.\ search all contiguous spans and calibrate the
false-positive rate for that search \cite{kirchenbauer2024reliability}. Both rely on an edit
disturbing tokenization only near the edit. Golowich and Moitra tolerate a constant fraction of
adversarial edits, but with a new construction built on pseudorandom codes
\cite{golowich2024edit}. A fixed vocabulary partition keeps the key synchronized under a positional
shift, which roughly doubles the provable edit tolerance relative to the soft watermark keyed on
the previous token without making the mark immune to edits \cite{zhao2024provable}. Watermarks on
sentence embeddings avoid tokenization but need an encoder and a natural segmentation into units
\cite{hou2024semstamp}, which a DNA verifier lacks.

Six-base tokens remove the locality these methods rely on. One inserted or deleted base shifts the
reading frame, so every later token changes identity, and so does the keyed context that sets each
mark bit. Alignment cannot repair this, because no token sequence survives to be aligned, and a
fixed partition does not help, because the changed tokens carry independent mark bits. What
survives is a contiguous region read in one of six phases. Verification therefore becomes a search
over strands, phases, start positions, and window lengths, and the correction for that search is
part of the method.

Four properties separate these constructions here: whether verification needs the generating
model; whether it needs an alignment, a known boundary, or a known strand; which edits it claims to
tolerate; and whether sampling is distorted. Biased-logit schemes distort sampling and are usually
verified with known boundaries \cite{kirchenbauer2023watermark}. Distribution-preserving schemes
realign to a key sequence \cite{kuditipudi2024robust} or certify a radius of token changes
\cite{wu2024dipmark}. Watermarks for biological sequence assume a coding region
\cite{zhang2025securing} or mark the protein, not its DNA \cite{chen2025protein}. The construction evaluated here needs no model, alignment, known boundary, or coding region, and it
preserves the sampling distribution in expectation.

\subsection{SynthID}

We use the binary tournament construction of SynthID-Text \cite{dathathri2024scalable}. At each
generation step, a secret key and the most recent tokens assign every candidate token a binary
mark bit in each tournament layer, and each layer favors candidates whose mark bit is 1. Averaged
over fresh keyed functions, the tournament leaves the next-token distribution unchanged. At a
fixed key and context, it samples from a reweighted distribution. The detector rebuilds the mark
bits from the output and tests for an excess of 1s. Because the bits depend on local context, not
on absolute position, they can be rebuilt from any retained stretch of sequence, and the detector
needs only the output, the key, and published detector settings
(Section~\ref{sec:detection}). Keying on recent tokens is common to
autoregressive text watermarks \cite{kirchenbauer2023watermark}. We chose SynthID because it
combines local keying with tournament sampling and has been evaluated in a large-scale text
deployment.

\subsection{Genomic tokenization and token phase}

Several genomic models, including both studied here, divide DNA into non-overlapping six-base
tokens, which shortens the token sequence and lets a model cover more bases in a fixed context
\cite{dallatorre2025nucleotide,wu2025generator,carbon2026,generatorv22026}. Models with one token
per base have no phase problem, because an insertion or deletion moves tokens without changing them
\cite{nguyen2024evo,nguyen2023hyenadna}. Caduceus uses one token per base because small input
changes can otherwise change the tokenization drastically \cite{schiff2024caduceus}. GENERator
randomizes the tokenization offset of each training sample \cite{wu2025generator}, and
GENERator-v2 cycles through all six offsets \cite{generatorv22026}. None of these remedies is
available to a verifier that receives only DNA, which must recover the phase from the sequence
itself.

\section{Methods}
\label{sec:methods}

\subsection{Models and sampling}

Generation used Carbon-500M \cite{carbon2026,carbon500m} and GENERator-v2-eukaryote-1.2b-base
\cite{generatorv22026,generatorv2checkpoint}, two decoder-only models. At every step, we restricted the logits to the 4{,}096 canonical
DNA tokens and normalized once. For GENERator, this is its direct token distribution, not the
per-base sampling that its released code applies by default. Temperature was 1.0, with no top-$k$
or top-$p$ truncation. The resulting distribution $p$ is shared by both samplers: the ordinary
sampler draws from $p$ without a key, and the marked sampler applies the tournament below. We
report each model separately.

\subsection{Watermark embedding}

At each step, the key, the preceding four tokens, and a candidate token determine the candidate's
mark bit $g$, computed with HMAC-SHA-256 over the key, a fixed domain label, the context, and the
candidate. The tournament formula, the number of layers, the context width, and the
repeated-context rule follow the public reference implementation of SynthID-Text
\cite{synthidtextcode}. One tournament layer reweights the distribution as
\begin{equation}
p'(x) \;=\; p(x)\,\Bigl(1 + g(x) - \textstyle\sum_y p(y)\, g(y)\Bigr),
\label{eq:layer}
\end{equation}
so candidates with a mark bit of 1 gain probability by an amount set by the total probability of
such candidates. The sampler applies $m = 30$ layers, each with its own mark bits, and samples from
the resulting distribution $p_k$ (Figure~\ref{fig:method}a).

The first four generated tokens are sampled ordinarily, so every later context can be rebuilt from
the continuation alone. A four-token context that has already occurred is also sampled ordinarily
and excluded from the detector score, as in the original construction
\cite{dathathri2024scalable}; the sampler keeps a history of 1{,}024 contexts to recognize
repeats. Averaged over fresh keyed functions, Equation~\ref{eq:layer} leaves the next-token
distribution unchanged. At a fixed key and context, the sampler follows the reweighted
distribution $p_k$, so the sampler checks test against $p_k$.

\subsection{Position-independent detection}
\label{sec:detection}

\begin{figure*}[tp]
\centering
\includegraphics[width=\textwidth]{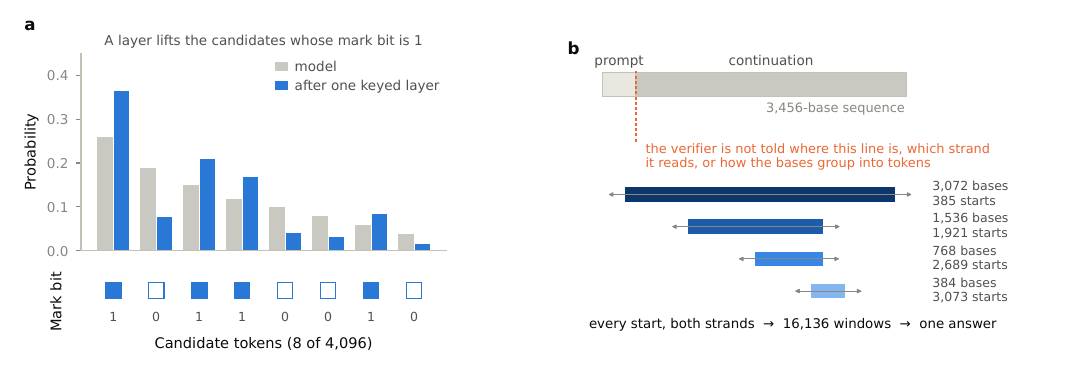}
\caption{\textbf{SynthID embedding and position-independent detection.}
\textbf{a}, One keyed layer, drawn for eight of the 4{,}096 candidate tokens with an illustrative
distribution. Candidates whose mark bit is 1 gain probability, and the others lose it. The sampler
applies 30 such layers. Panel \textbf{a} illustrates Equation~\ref{eq:layer} and is not a
measurement.
\textbf{b}, Detector inputs and search. The verifier receives the sequence, the key, and the published detector
settings, without the generation boundary, the strand, or the token phase. It scores windows of
four lengths at every starting base on both strands, 16{,}136 windows for a 3{,}456-base sequence, and
reports one sequence-level result corrected for the full search.}
\label{fig:method}
\end{figure*}

The verifier receives a DNA sequence, the key, and the published detector settings: the
generation domain label, the four window lengths, the 30 tournament layers and four-token context
used to rebuild the mark bits, the 1{,}024-context repetition history, and the significance level
$\alpha = 0.01$. In deployment, only the key would be secret.
It does not receive the prompt, the model or its probabilities, the seed, the strand, the token
phase, or the generation boundary (Figure~\ref{fig:method}b).

The verifier treats every candidate generated region as a window: a stretch of the sequence on one
strand, starting at a given base, with one of four lengths. It scores windows at every starting
base of the sequence and of its reverse complement, which covers every start position and all six
token phases. Inside a window, the DNA is divided into tokens from the window's first base, and the
first four tokens supply context. Every later token whose context is not a repeat is a scored
token. The verifier rebuilds the mark bits of the scored tokens and counts the ones, $S$, among
their $n$ mark bits.

If the sequence carries no mark for this key, the mark bits behave as independent fair coin flips, and
the window $P$-value is the binomial tail
\begin{equation}
P_{\mathrm{win}} \;=\; \Pr\bigl[\mathrm{Binomial}(n, \tfrac{1}{2}) \geq S\bigr].
\end{equation}
For a sequence of $N$ bases, the verifier scores
\begin{equation}
M \;=\; 2\sum_{L} (N - L + 1), \qquad L \in \{384,\, 768,\, 1{,}536,\, 3{,}072\},
\end{equation}
windows, which is 16{,}136 for a 3{,}456-base sequence, and makes one decision for the sequence:
\begin{equation}
\begin{aligned}
P_{\mathrm{seq}} &\;=\; \min\bigl(1,\; M \cdot \min_{\text{windows}} P_{\mathrm{win}}\bigr),\\
\text{mark found} &\iff P_{\mathrm{seq}} \leq \alpha .
\end{aligned}
\end{equation}
The Bonferroni factor $M$ covers every window searched, including overlapping ones, and the level
$\alpha$ was set in advance. We call $-\log_{10} P_{\mathrm{win}}$ the strength of a window.
A sequence is positive exactly when some window reaches strength $\log_{10}(M/\alpha) = 6.21$. The
search depends only on the sequence, so it is the same for both models, and $M$ is recomputed when an
edit changes the sequence length. Supplementary Section~\ref{supp:cost} gives the cost of the search.

Bonferroni correction needs no independence between tests, but it is conservative here, because
windows of the same length whose starts differ by a multiple of six bases share a token phase and
nearly all their tokens (Supplementary Section~\ref{supp:bonferroni}).

A window need not be intact to be detected. A token that has lost its phase or its context
contributes unbiased bits and only dilutes the count, so a window is positive once it holds enough
tokens that kept both. In the shortest window, about 56\% of the mark bits must be 1
(Supplementary Section~\ref{supp:geometry}).

\subsection{Prompts and cohorts}
\label{sec:prompts}

Prompts are 384-base forward-strand windows from 16 accession-versioned RefSeq chromosome records
of eight model organisms. Windows were chosen by hashing a fixed label, the accession, and a
segment index. Human, patient, controlled-access, organelle, plasmid, and ambiguous-base spans
were excluded. Each cohort was fixed before any model generated from it, and no prompt was added,
moved, or removed afterward.

The development cohort contains 256 prompts. Before any detection result was examined, a hash rule
reserved 64 of them for choosing how to summarize the mark bits in a window (Supplementary
Section~\ref{supp:selection}) and held out the other 192. The sampler check and the sequence
measures use this cohort only.

The detection cohort contains 1{,}608 prompts, 134 non-overlapping windows from each of 12 of the
16 records, covering six of the eight organisms (not mouse or budding yeast), drawn with a separate
selection label. The same hash rule set aside 64 prompts, which were not used, and held out
1{,}544. A separate generation domain label gave both models new keyed functions, and all other
settings matched. The same authors generated and scored this cohort with the same code, in a
separate computing environment.

Every prompt produced two marked and two ordinary continuations of 512 tokens (3{,}072 bases) per
model. The two draws used different seeds, and the two marked draws used different keys, both
published for reproducibility. Prepending the prompt to a held-out continuation gives the
3{,}456-base test sequence that the verifier receives, either a marked sequence or an ordinary
sequence; its generation boundary is withheld. Scoring each marked sequence with the other marked draw's key gives the wrong-key control,
which tests whether a positive depends on the correct key. The three rates are reported
separately, because a pooled accuracy would depend on how common marked sequences are in an
application.

\subsection{Sampler validation}
\label{sec:sampler-validation}

The correctness of the marked sampler rests on implementation tests in the project's test suite.
The tournament update reproduces the public SynthID-Text reference implementation
\cite{synthidtextcode} on peaked, flat, random, and degenerate distributions. The vectorized
computation of $p_k$ used during generation agrees with an independent unvectorized implementation
to floating-point precision. Computing $p_k$ once and applying its inverse-CDF sampling step
reproduces the generation sampler's draws exactly.

A second check uses realistic model states. For each model and development prompt, we fixed the
state at token index 64 of the first ordinary continuation and drew 5{,}000 samples from each of
$p$ and $p_k$. Each set of counts was compared with its source distribution by the likelihood-ratio
($G$) statistic. An expected count of about 1.2 per category is too low for the asymptotic
$\chi^2$ reference, so $G$ was referred to a Monte Carlo null of 999 samples from the same
distribution, which is exact by construction. Because the draws come from the computed
distributions, this check confirms the sampling step and the test, not the computation of $p_k$.
The 256 tests per sampler form one family under Benjamini--Hochberg correction at 0.05. Bonferroni
correction cannot reject here, because 999 replicates allow no $P$-value below $10^{-3}$, above its
level of $1.95 \times 10^{-4}$. As a negative control, at eight further states per model, draws
from $p$ were tested against $p_k$.

\subsection{Sequence measures}

The model score is the teacher-forced negative log-likelihood per generated token under the same
model and token restriction as generation, with the prompt as context only. Each model scores its
own outputs, so scores are not compared between models. Thirteen further measures were specified
before analysis: GC content; purine content; CpG content; base, dinucleotide, and trinucleotide
entropy; distinct 6-mer fraction; longest and mean single-base run; Jensen--Shannon drift from the
prompt at $k = 1, 2, 3$; and perplexity. All 14 measures form one family under
Benjamini--Hochberg correction.

\subsection{Edits and statistics}

In the single-edit conditions, each sequence has exactly one substituted, inserted, or deleted base in
the continuation, placed by a fixed pseudorandom rule that never consults the verifier. Prompts are
the independent unit, and a prompt's two draws are paired repetitions. Uncertainty in the sequence
measures used an equal-weight bootstrap over prompts with 20{,}000 replicates and a paired
sign-flip test with 100{,}000 replicates. A control prompt is positive if either draw is positive;
for marked prompts, we also require both draws to be detected.

The detection cohort's analysis plan, specified before generation, reports the exact one-sided
95\% upper confidence bound (Clopper--Pearson) on the false-positive rate over the 1{,}544
held-out prompts. Ordinary sequences, the false positives that matter in use, are the primary control,
and the wrong-key control is a secondary check of key specificity. The prompt-level rate is about
twice the per-sequence rate that $\alpha$ controls, so comparing it with the 1\% target is
conservative. Development-cohort rates carry exact two-sided 95\% intervals.

\subsection{Edit-rate experiment}
\label{sec:edit-rate-methods}

To find where random editing defeats the verifier, we rescored the detection cohort's stored sequences
after applying $k = \mathrm{round}(3{,}072\,r)$ edits to each continuation. We call $r$ the
per-base edit rate and tested seven values from 0.03\% to 10\% (Figure~\ref{fig:editrate}). A
substitution changes one base. Each insertion or deletion adds or removes one to five bases, so the
edit rate counts edit events, not changed bases. Four series were run: substitutions, insertions
only, deletions only, and a mixed series of random insertions and deletions. The single-direction
series keep opposite frame shifts from cancelling. Positions and indel lengths came from a public
hash stream set by the series, the sequence, and the rate, never by the detector, and no sequence or
detector setting changed. Marked sequences were scored with the correct key and ordinary sequences with the
corresponding key, 3{,}088 sequences per model, rate, and series, without wrong-key controls. The
analysis plan required the ordinary-sequence rate to stay at or below 1\% in every cell, because a
rise with edit rate would mean that the edits themselves create apparent signal.

\section{Results}
\label{sec:results}

\subsection{Sampler fidelity}

At the tested model states, draws from the fixed-key and model distributions agreed with those
distributions, and the test could tell the two distributions apart. Each model had 256 tested
states, and chance alone predicts 12.8 of them to reject at $P < 0.05$. The marked sampler produced 13 in
Carbon and 5 in GENERator, and the ordinary sampler produced 12 and 14. In GENERator, neither
sampler had a rejection after Benjamini--Hochberg correction. For Carbon, the corrected count was
not retained with the evidence, and its nominal counts are close to chance. The negative control,
draws from the model's distribution tested against the fixed-key distribution, was rejected at all
eight control states in each model. Together with the implementation tests
(Section~\ref{sec:sampler-validation}), these results support the correctness of the sampler at
the resolution of 5{,}000 draws per state. They do not prove it correct at every possible state.

\subsection{Sequence-quality proxies}

\begin{figure*}[tp]
\centering
\makebox[\textwidth][l]{\small\textbf{Carbon-500M}}\par\vspace{0.15em}
\includegraphics[width=\textwidth]{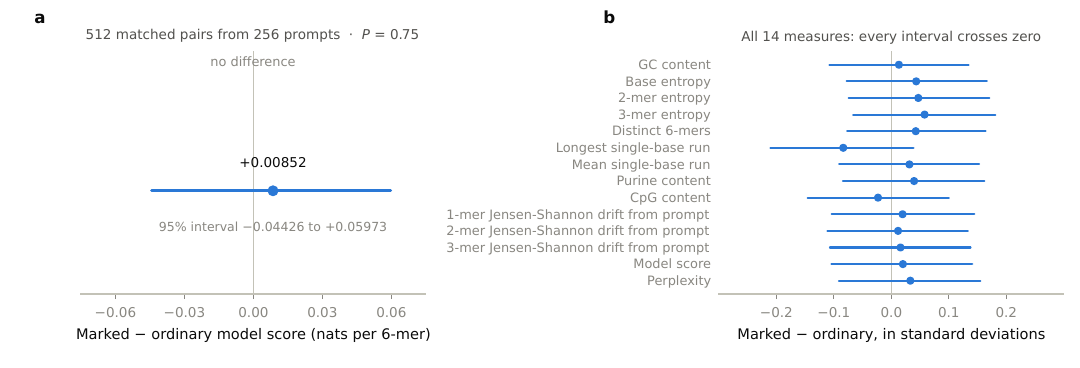}
\vspace{0.35em}
\makebox[\textwidth][l]{\small\textbf{GENERator-v2 1.2B}}\par\vspace{0.15em}
\includegraphics[width=\textwidth]{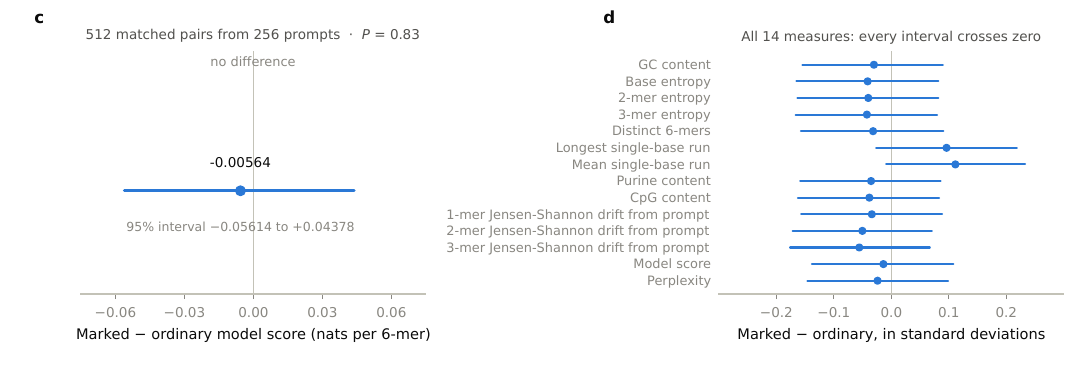}
\caption{\textbf{Model likelihood and sequence-quality proxies in the development cohort.}
\textbf{a,c}, Paired difference in model score between marked and ordinary continuations, over 512
matched pairs from 256 prompts. Positive values indicate lower model likelihood for marked
continuations. The horizontal bar is the 95\% interval from a bootstrap over prompts: $-0.04426$ to
$+0.05973$ in Carbon ($P = 0.75$) and $-0.05614$ to $+0.04378$ in GENERator ($P = 0.83$).
\textbf{b,d}, Paired differences for all 14 measures in standard deviations, with intervals in the
same units. Every interval includes zero, and none of the 14 comparisons is significant after
multiple-testing correction. Panels \textbf{a,b} show Carbon and panels \textbf{c,d} show
GENERator.}
\label{fig:quality}
\end{figure*}

Watermarking produced no statistically detectable difference between marked and ordinary
continuations in either model (Figure~\ref{fig:quality}). Every interval for the 14 measures
contained zero, and no comparison was significant after correction. The smallest adjusted
$P$-value was 0.85 in Carbon and 0.73 in GENERator. The paired difference in model score was
$+0.00852$ nats per token in Carbon and $-0.00564$ in GENERator. The largest absolute standardized
effects were 0.084 standard deviations in Carbon, on the longest single-base run, and 0.111 in
GENERator, on the mean single-base run. These results indicate no measurable quality loss under the
evaluated proxies.

\subsection{Detection and false-positive rate}
\label{sec:second-cohort}

\begin{figure*}[tp]
\centering
\makebox[\textwidth][l]{\small\textbf{Carbon-500M}}\par\vspace{0.15em}
\includegraphics[width=\textwidth]{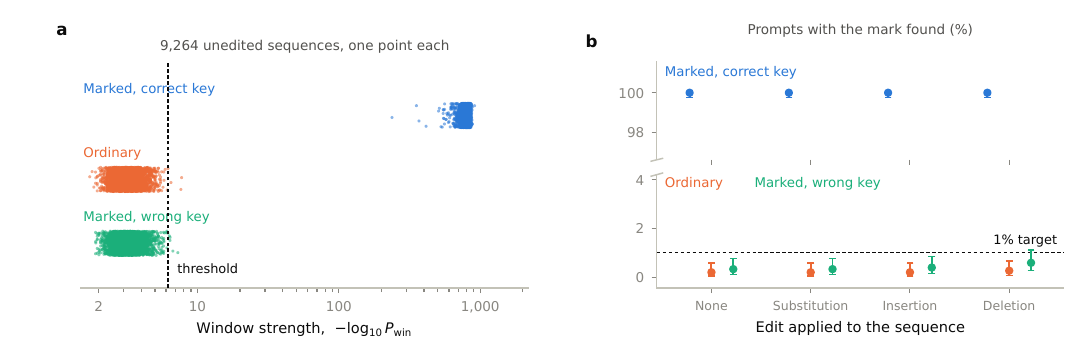}
\vspace{0.35em}
\makebox[\textwidth][l]{\small\textbf{GENERator-v2 1.2B}}\par\vspace{0.15em}
\includegraphics[width=\textwidth]{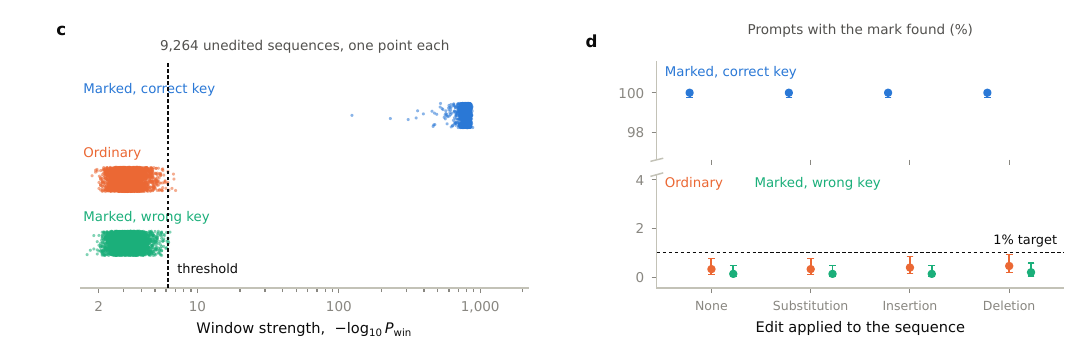}
\caption{\textbf{Position-independent detection in the detection cohort.} \textbf{a,c}, Strength of
the best window in each unedited sequence, one point per sequence, on a log scale: 3{,}088 marked sequences
scored with the correct key, 3{,}088 ordinary sequences, and 3{,}088 wrong-key controls per model. The
dashed line is the threshold corrected for the search over 16{,}136 windows. All marked sequences
exceed it. \textbf{b,d}, Proportion of the 1{,}544 held-out prompts with positive detections (both
draws for marked sequences with the correct key, either draw for controls), with exact two-sided 95\%
intervals, on a broken axis. All marked prompts were detected in all four edit conditions.
Table~\ref{tab:second} gives the counts and the one-sided upper bounds used for the false-positive
comparison, and Supplementary Figure~\ref{fig:primary-detection} shows the development cohort.
Panels \textbf{a,b} show Carbon and panels \textbf{c,d} show GENERator.}
\label{fig:detection}
\end{figure*}

In the detection cohort, the verifier found every marked sequence in both models, unedited and after
each single-base edit (Table~\ref{tab:second}; Figure~\ref{fig:detection}). That is 3{,}088 sequences
per model and condition, and it places the exact two-sided 95\% interval for the prompt-level
detection rate at 99.76--100\%.

Every one-sided 95\% upper confidence bound on the false-positive rate for ordinary sequences lay below
the 1\% target. The largest was 0.850\%, for GENERator after a deletion. The analysis plan had
stated before generation that a bound below 1\% required at most eight positive prompts.

The wrong-key control, a secondary check of key specificity, crossed 1\% in one cell. For Carbon
after a deletion, its bound was 1.015\%.

The correction was conservative in practice. Per sequence, the corrected decision was positive for
0.105\% of ordinary Carbon sequences and 0.186\% of ordinary GENERator sequences, pooled across the four
conditions, well below $\alpha$ (Supplementary Section~\ref{supp:bonferroni}).

The development cohort gave the same result. All 384 marked sequences per model were found in every
condition, and each control produced at most one positive sequence per model and condition
(Supplementary Table~\ref{tab:detection}).

Typical detection strength was the same in both models. The median unedited marked sequence reached
strength 792.9 in Carbon and 799.7 in GENERator, far above the threshold. The weakest detected sequences
differ more, but in the development cohort that difference came from a single GENERator sequence
with a highly repetitive continuation (Supplementary Section~\ref{supp:weakest}).

\begin{table*}[tp]
\centering
\caption{Positive detections in the detection cohort, for each model across 1{,}544 held-out prompts
with two draws each (3{,}088 sequences per model and condition). Correct-key entries count marked
sequences. Control entries count positive sequences, each from a different prompt; the value in
parentheses is the exact one-sided 95\% upper confidence bound on the prompt-level rate, in
percent. Ordinary sequences were designated in advance as the primary false-positive measure, and the
wrong-key control is a secondary check of key specificity. Each edit changes exactly one base.}
\label{tab:second}
\small
\setlength{\tabcolsep}{4pt}
\renewcommand{\arraystretch}{1.08}
\begin{tabular}{@{}lccc@{\hspace{12pt}}ccc@{}}
\toprule
& \multicolumn{3}{c}{\textbf{Carbon-500M}} & \multicolumn{3}{c}{\textbf{GENERator-v2 1.2B}} \\
\cmidrule(lr){2-4}\cmidrule(l){5-7}
Edit applied to the sequence & Correct key & Ordinary & Wrong key & Correct key & Ordinary & Wrong key \\
\midrule
None             & 3{,}088/3{,}088 & 3 (0.501) & 5 (0.680) & 3{,}088/3{,}088 & 5 (0.680) & 2 (0.407) \\
One substitution & 3{,}088/3{,}088 & 3 (0.501) & 5 (0.680) & 3{,}088/3{,}088 & 5 (0.680) & 2 (0.407) \\
One insertion    & 3{,}088/3{,}088 & 3 (0.501) & 6 (0.766) & 3{,}088/3{,}088 & 6 (0.766) & 2 (0.407) \\
One deletion     & 3{,}088/3{,}088 & 4 (0.592) & 9 (1.015) & 3{,}088/3{,}088 & 7 (0.850) & 3 (0.501) \\
\bottomrule
\end{tabular}
\end{table*}

\subsection{Detection after single-base edits}
\label{sec:single-edits}

\begin{figure*}[tp]
\centering
\makebox[\textwidth][l]{\small\textbf{Carbon-500M}}\par\vspace{0.15em}
\includegraphics[width=\textwidth]{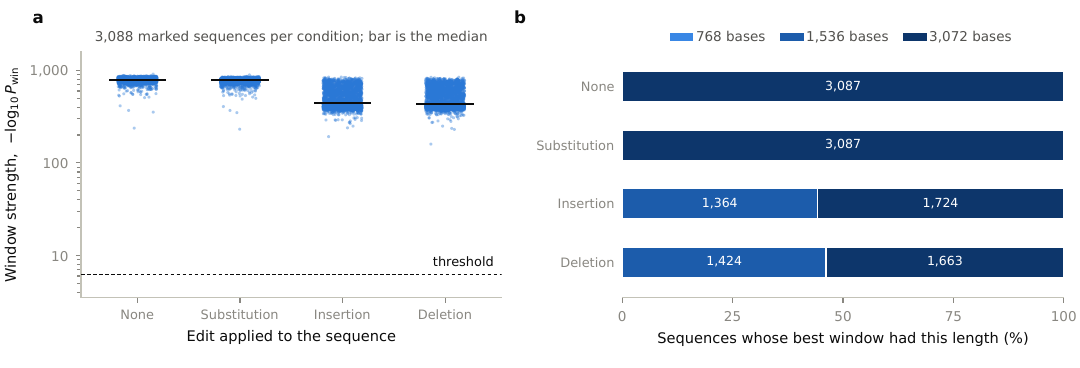}
\vspace{0.35em}
\makebox[\textwidth][l]{\small\textbf{GENERator-v2 1.2B}}\par\vspace{0.15em}
\includegraphics[width=\textwidth]{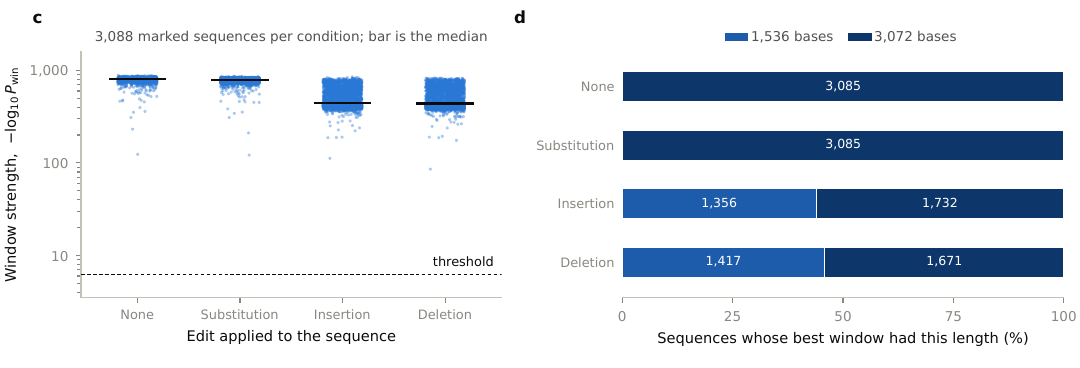}
\caption{\textbf{Effects of single-base edits on detection in the detection cohort.} \textbf{a,c},
Strength of the best window in each of the 3{,}088 marked sequences per condition, on a log scale, with
the median shown as a bar. The dashed line is the detection threshold. \textbf{b,d}, Length of the
strongest window; numbers inside the bars are sequence counts out of 3{,}088. The full 3{,}072-base
window is strongest for nearly all unedited and substituted sequences, whereas shorter windows are
selected for almost half of the sequences after an insertion or deletion. Supplementary
Figure~\ref{fig:primary-edits} shows the development cohort. Panels \textbf{a,b} show Carbon and
panels \textbf{c,d} show GENERator.}
\label{fig:edits}
\end{figure*}

One inserted or deleted base cost a sequence almost half of its strength, and one substituted base cost
little, yet every marked sequence stayed detectable (Figure~\ref{fig:edits}a,c). A substitution lowered
the median strength by about 2\%, and an insertion or deletion by about 45\% (Supplementary
Table~\ref{tab:window}).

An insertion or deletion also changed which window carried the evidence. For unedited and
substituted sequences, the full 3{,}072-base window was strongest in nearly every sequence. After an
insertion or deletion, a shorter window was strongest in almost half of the sequences
(Figure~\ref{fig:edits}b,d). A substitution changes token content and nearby contexts but keeps the
downstream token phase. An insertion or deletion shifts the division of every later token in a
window that spans the edit. The shift toward shorter windows is consistent with the search
selecting regions whose token phase is preserved or recovered. The development cohort showed the
same shift (Supplementary Figure~\ref{fig:primary-edits}).

\begin{figure*}[tp]
\centering
\includegraphics[width=\textwidth]{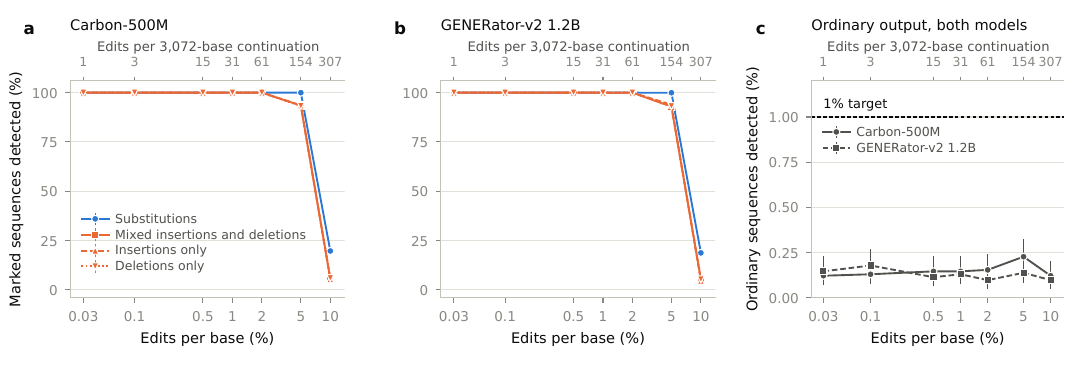}
\caption{\textbf{Detection as random edits accumulate, detection cohort.} \textbf{a,b}, Percentage
of 3{,}088 marked sequences detected with the correct key in Carbon (\textbf{a}) and GENERator
(\textbf{b}), by per-base edit rate (log scale) and edit type. The rates were 0.03, 0.1, 0.5, 1, 2,
5, and 10\%; the top axis gives the corresponding 1, 3, 15, 31, 61, 154, and 307 edits per
3{,}072-base continuation. Each insertion or deletion adds or removes one to five bases, and in the
mixed series each edit is an insertion or a deletion at random. \textbf{c}, Percentage of ordinary
sequences detected under the corresponding key, pooled over the four edit types (12{,}352 sequences per
point), against the 1\% target. Pooled over every rate and edit type, including the unedited sequences,
132 of 89{,}552 Carbon sequences and 116 of 89{,}552 GENERator sequences were positive. Vertical bars are
exact two-sided 95\% intervals and are often smaller than the markers. Edit positions were chosen
without reference to the detector.}
\label{fig:editrate}
\end{figure*}

\subsection{Detection as edits accumulate}
\label{sec:edit-rate}

Detection stayed complete or nearly complete through a 2\% per-base edit rate, that is, 61 edits
per continuation, in both models (Figure~\ref{fig:editrate}). Every marked sequence was found at every
rate up to 2\% for every edit type, except for one GENERator sequence in the mixed series at 2\%.

Insertions and deletions defeated the verifier at lower rates than substitutions. At a 5\% rate,
substitutions were still detected in every Carbon sequence and all but one GENERator sequence, whereas
detection after insertions and deletions fell to 92.9--93.9\%. At 10\%, substitutions were
detected in 19.7\% of Carbon sequences and 18.7\% of GENERator sequences. The other edit types fell to
4.6--5.9\%.

The two models' curves agreed within one percentage point at every rate and edit type, so the
single GENERator miss at 2\% does not indicate a difference in robustness. The insertion-only and
deletion-only series behaved like the mixed series, so cancelling frame shifts do not explain the
tolerance.

Ordinary sequences stayed below the 1\% target at every rate and in every series. Pooled over all
rates and edit types, 0.15\% of ordinary Carbon sequences and 0.13\% of ordinary GENERator sequences were
positive. The largest value in any single cell was 0.29\%, for GENERator in the mixed series at
1\%.

\section{Discussion}

A verifier that knows neither the strand, the start of the generated region, nor the token phase
can recover a generation-time watermark from DNA, with one decision corrected for the whole search.
This shared watermarking and detection procedure is the methodological contribution.

The verifier tolerates random edits because it does not need an intact window. A simple argument
predicts failure much earlier: at a 1\% rate, stretches free of insertions and deletions average
about 100 bases, shorter than the 384-base minimum window. A window, however, gains excess 1s from
every token that keeps its phase and context, and the other tokens only dilute the count. An
unedited sequence starts near strength 800 against a threshold of 6.21, so it can lose most of its
evidence before the decision changes. The shift toward shorter windows after an insertion or
deletion fits the same picture, with the search finding the original phase again downstream of the
edit. The strongest window does not identify the exact generation boundaries.

Running the same sampler and verifier in two models shows that the implementation carries across
them. The two models agreed in detection, had nearly identical typical strength, and showed no
measurable quality loss, although they share much of their training data.

The margin seen here should not be assumed for shorter generated regions or longer sequences. A
shorter region supplies fewer mark bits, and a longer sequence adds windows to the correction. The
shortest detectable continuation, and the trade-off between detection power and distortion across
tournament depths, remain to be measured in both models.

The search addresses a synchronization problem that coding theory treats directly.
Synchronization strings add indexing information to recover from insertions and deletions
\cite{haeupler2017synchronization}, and pseudorandom codes combine indistinguishability with
decoding under specified errors \cite{christ2024prc}. Davey and MacKay's watermark codes integrate
over the accumulated drift with a hidden Markov model, but they presume control of the encoding,
which a sampling watermark does not have \cite{davey2001reliable}. Our verifier adds no indexing
code to the sequence. It enumerates the offsets, pays for the search through its correction, and
carries none of the error-correction guarantees of those codes. Synonymous-codon watermarks for
coding sequences \cite{zhang2025securing} serve a complementary goal, and combining them with
generation-time marks would need its own evaluation.

\section{Limitations}
\label{sec:limitations}

This evaluation covers two models, one watermark construction, and one verifier, and it gives no
general proof of watermark robustness, biological safety, or secret-key security. Both models use
six-base tokens; single-base or variable-length tokens would pose different alignment problems.
Both cohorts come from the same 16 chromosome records or a subset of them, which limits
extrapolation to other genomic sources, and longer sequences would increase the search correction. The
models are not independent, because Carbon's training corpus is built mainly on GENERator-v2's
pre-training data, and their agreement does not make them equivalent in any respect.

We compared SynthID with its matched ordinary sampler, not with another watermark construction on
the same cohort, such as a biased-logit scheme adapted to six-base tokens. That comparison would
confound the construction with the verifier that is our contribution, and the property comparison
in Section~\ref{sec:background} does not substitute for measured performance on a shared cohort.

The sampler and quality results come from one execution per model on the development cohort. The
detection cohort was generated and scored in a separate computing environment with new keyed
functions, but by the same authors with the same code, so it is not an independent replication.
For Carbon, the protocol document retained with the development-cohort detection result does not
match the hash recorded in that result, and the original document could not be recovered. An audit
found that eight of the nine settings it governs are recorded in the result's own fields and agree
with the reported analysis. The ninth, the 1{,}024-context repetition history, rests only on the
default of the detector code identified by its recorded hash. The mismatch therefore gives no
indication that the sequences or the detection statistic differ from what we report, but the
execution cannot be checked against the document itself.

The false-positive evidence is narrow. With 192 prompts, the development cohort could not show a
rate below 1\% (Supplementary Table~\ref{tab:detection}). In the detection cohort, every bound for
ordinary sequences lies below 1\%, but the largest, 0.850\%, does so narrowly, and the wrong-key bound
for Carbon after a deletion exceeds that level at 1.015\%. These rates apply to the sequence length and
sequence sources tested here; elsewhere, false-positive control rests on the Bonferroni correction
alone.

Every edit was placed at random, without reference to the detector, and the edit-rate series scored
no wrong-key controls. We did not test an adaptive editor who inspects the detector, the key, or
any score and places edits where they do the most damage.

The keys are published for reproducibility, so secrecy, resistance to key recovery, and removal in
deployment were not evaluated, and the analytic null assumes independent fair mark bits without
the correct key. Attacks reported against text watermarks remain untested here, including key
recovery from the detector's $P$-values \cite{reynolds2025breaking}, paraphrasing and
back-translation of SynthID-Text \cite{han2025synthid}, and a layer-inflation attack shown to
affect the mean score that our verifier also uses \cite{omidi2026synthid}.

Model likelihood and sequence statistics are proxies, not measures of expression, replication,
function, viability, or safety, and our quality claim covers only the 14 measures. Standardized
in-silico suites, such as supervised downstream tasks and zero-shot variant-effect benchmarks
\cite{marin2024bend,liu2024genbench,patel2024darteval}, score a model's reading of annotated DNA
and mostly rely on labels tied to genome coordinates that a new generation does not carry.
Evaluating marked outputs against them would need a paired design or an experimental assay, which
is the natural next step before any claim about biological consequence.

\section{Conclusion}

The SynthID tournament watermark can be embedded in Carbon-500M and GENERator-v2 1.2B without a
statistically detectable change in the evaluated quality proxies. A verifier holding the key can
recover it from DNA by searching strands, start positions, and window lengths under one
correction. In the detection cohort, the verifier found every marked sequence in both models, unedited
and after one edited base, and the one-sided 95\% upper confidence bound on the false-positive rate
for ordinary sequences was at most 0.850\%. Detection survived random edits up to a 2\% per-base rate
and failed for most sequences at 10\%. Broader conclusions require independent replication, wider
sequence coverage, tests against adaptive editing, and biological validation.

\section*{AI use statement}

Generative AI tools assisted with method implementation, experiment design, statistical
derivations, data preparation, literature retrieval, and interpretation of results. They also
assisted with drafting and revising the manuscript, articulating the verification setting and
claim scope, and preparing the \LaTeX{} submission. Carbon-500M and GENERator-v2 1.2B generated
the DNA continuations studied in the experiments, as described in Section~\ref{sec:methods}.
The paper contains no formal theorems or proofs. Its statistical development uses standard
probability models and a Bonferroni correction, and the authors verified all derivations,
numbers, and claims against the recorded measurements. The authors are responsible for the final
text, claims, and artifacts.

\section*{Ethics statement}

The prompt cohorts use public non-human RefSeq records and exclude patient and
controlled-access sequences (Section~\ref{sec:prompts}). The experiments assess statistical
detection and sequence proxies. They do not establish biological function, viability, safety, or
sequence authenticity. The experimental keys are published for reproducibility and provide no
evidence of secrecy or resistance to key recovery. A detected watermark is evidence under the
specified statistical model, not a certificate of biological origin or safe use.

\section*{Reproducibility statement}

Section~\ref{sec:methods} specifies the canonical token policy, the tournament update, the
detector search and its correction, prompt construction, and the statistical comparisons, and the
Supplementary information gives the window-score selection, the search cost, and the computing
environment. The figures and tables report both models separately under the same evaluation
design. Section~\ref{sec:limitations} records the execution scope, the absence of an independent
replication, and the Carbon protocol-document discrepancy together with its audit. The code,
protocol documents, and evidence ledger needed to reproduce the analysis are available as
described in the Data and code availability statement.

\section*{Data and code availability}

The verifier implementation, the watermark sampler, the prompt-cohort construction scripts, the
analysis and figure scripts, the protocol documents, and the evidence ledger to which every number
in this paper resolves are maintained in the
\href{https://github.com/Georgakopoulos-Soares-lab/genomic-sampling-watermarks}{project GitHub
repository}. The repository includes the detection results for both models in the detection
cohort and the GENERator results for the development cohort. The Carbon development-cohort result
files, generated sequences, and other underlying files not in the repository are available from
the authors on request. The sampler's implementation tests, including the comparison with the
SynthID-Text reference implementation at the cited commit, are part of the repository's test
suite. The two experimental keys are published for reproducibility and are not deployment secrets.

\section*{Competing interests}

The authors declare no competing financial or non-financial interests.

\section*{Acknowledgments}

This work has been supported by the National Institute of General Medical Sciences of the National
Institutes of Health [R35GM155468 to I.G.S.]; and start-up funds awarded to I.G.S.

\begingroup
\small
\raggedright
\balance
\bibliographystyle{plain}
\bibliography{refs}
\endgroup

\clearpage
\section*{Supplementary information}
\setcounter{subsection}{0}
\renewcommand{\thesubsection}{S\arabic{subsection}}
\renewcommand{\theHsubsection}{S\arabic{subsection}}
\setcounter{figure}{0}
\renewcommand{\thefigure}{S\arabic{figure}}
\renewcommand{\theHfigure}{S\arabic{figure}}
\setcounter{table}{0}
\renewcommand{\thetable}{S\arabic{table}}
\renewcommand{\theHtable}{S\arabic{table}}

\subsection{Window-score selection}
\label{supp:selection}

Before any detection result was examined, 64 development prompts were reserved for choosing how to
summarize the mark bits in a window. Neither they nor the 192 held-out prompts were used to fit the
detection threshold. The two candidates were the released SynthID mean score over scored tokens and
the upstream default weighted-mean score. For GENERator, the two were compared on the 64 reserved
prompts alone. The comparison used continuations of 384 and 3{,}072 bases, scored with the token
phase known, and measured the separation between marked and ordinary continuations in units of the
ordinary-continuation standard deviation. The mean score separated them at least as well at both
lengths and was selected. No held-out prompt was scored during this selection, and no
signal-matched weights were fitted. We applied the same mean score to Carbon, but the equivalent
Carbon comparison is recorded only in the run notes, not as a retained result, so this selection is
independently verified for GENERator only.

\subsection{Window geometry}
\label{supp:geometry}

Each scored token carries 30 mark bits, one per tournament layer. The number of bits does not
limit detection: 21 bits that are all 1 already reach the corrected threshold, so one scored token
would suffice. What limits detection is dilution. The shortest window, 384 bases, holds 64 tokens.
The first four supply unwatermarked context, which leaves at most 60 scored tokens and at most
1{,}800 mark bits, fewer where a four-token context repeats. With all 60 scored, at least 1{,}004 of
the 1{,}800 bits must be 1, about 56\%. Longer windows need a smaller fraction. Detection therefore
depends on recoverable synchronization, not on the absence of edits.

\subsection{Conservatism of the correction}
\label{supp:bonferroni}

Bonferroni correction bounds the family-wise error rate without requiring independence between
tests. Because the searched windows are positively correlated, the effective number of independent
tests is smaller than $M$. The realized false-positive rate can then lie well below $\alpha$, and
recall is attained under a stricter threshold than necessary. The realized per-sequence rates in the
detection cohort (Section~\ref{sec:second-cohort}) are 9.5 times below $\alpha$ for Carbon and 5.4
times below for GENERator. The observed minimum window $P$-values behave like those of about
828--852 independent tests, consistently across both models and all four conditions. That is about
19 times fewer than the 16{,}136 windows searched. This moment-matched figure summarizes how
conservative the correction is. It does not replace the decision rule, which is unchanged.

\subsection{Search cost and computing environment}
\label{supp:cost}

Keyed hashing, not window enumeration, dominates the cost of the search. A sequence of $N$ bases has
twelve token frames, two orientations by six phases, and the mark bits of every token position in
a frame are computed once, in a single pass. Window statistics then follow from prefix sums over
those per-token bit counts, so each window of a given length is evaluated in constant time. Where a
frame contains repeated four-token contexts, the first-occurrence rule is enforced with Fenwick
trees, and each window costs $O(\log N)$ instead. The work is therefore $O(N)$ keyed hash
evaluations and $O(N \log N)$ additional time, with $O(N)$ space per sequence, and $M$ grows linearly in
$N$ for a fixed window set. Sequences shorter than every window length are not scored,
reverse-complement coordinates are mapped back onto the supplied sequence, and log-probabilities are
kept throughout to avoid floating-point underflow.

Timing was measured on an AMD EPYC 7763. One 3{,}456-base sequence took a median 0.0956 s on a single
thread. Sequences of 6{,}912 bases took 0.245 s, and sequences of 13{,}824 bases took 0.902 s. Peak memory
per process was 27.9 MiB for a 3{,}456-base sequence. For a 13{,}824-base sequence, it was 69.1 MiB.
Running 16 workers in parallel across sequences gave one sequence per 0.0167 s of wall-clock time. That is
5.7 times the single-threaded rate.

Development-cohort generation ran on GPUs. Model scoring ran on CPU for Carbon and on GPU for
GENERator. The detection cohort was generated on NVIDIA A100 GPUs in a separate computing
environment. Detection ran on x86-64 CPUs for both cohorts.

\subsection{Development-cohort detection}
\label{supp:development}

\begin{figure*}[tp]
\centering
\makebox[\textwidth][l]{\small\textbf{Carbon-500M}}\par\vspace{0.15em}
\includegraphics[width=\textwidth]{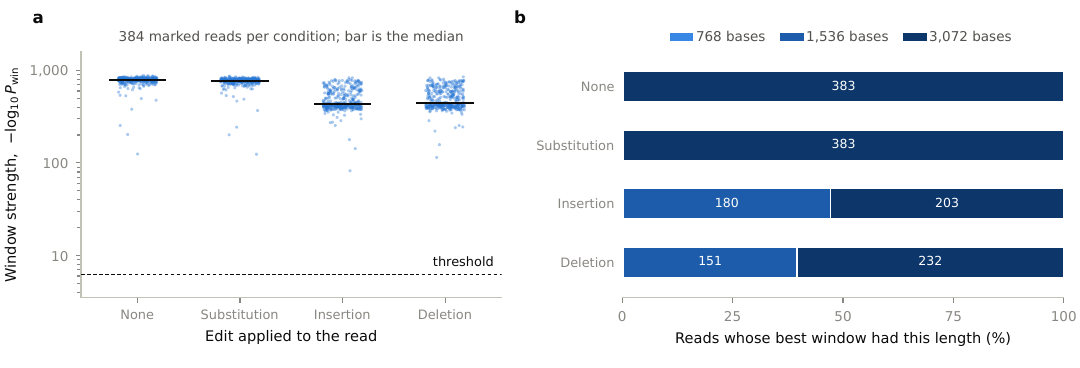}
\vspace{0.35em}
\makebox[\textwidth][l]{\small\textbf{GENERator-v2 1.2B}}\par\vspace{0.15em}
\includegraphics[width=\textwidth]{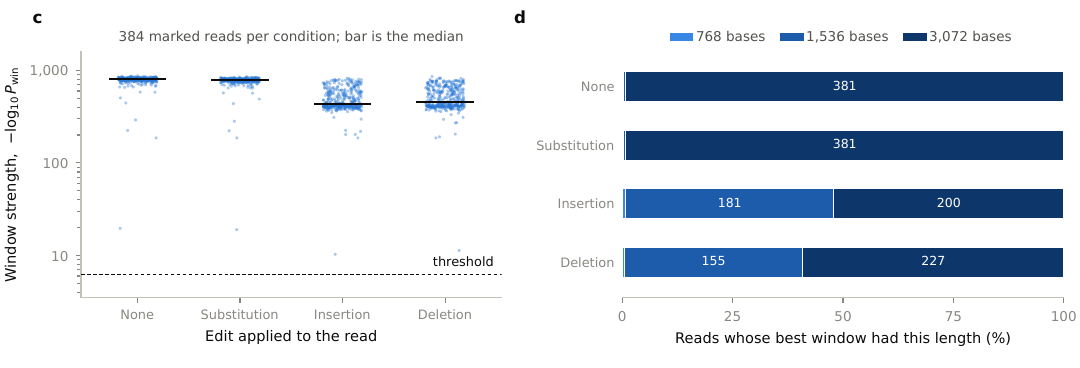}
\caption{\textbf{Single-base edits in the development cohort.} In the panel labels, ``reads''
denotes test sequences. \textbf{a,c}, Strength of the best
window in each of the 384 marked sequences per condition, on a log scale, with the median shown as a
bar. The dashed line is the detection threshold. \textbf{b,d}, Length of the strongest window;
numbers inside the bars are sequence counts out of 384. As in the detection cohort
(Figure~\ref{fig:edits}), the full 3{,}072-base window is strongest for nearly all unedited and
substituted sequences, whereas shorter windows are more frequent after an insertion or deletion. Panels
\textbf{a,b} show Carbon and panels \textbf{c,d} show GENERator.}
\label{fig:primary-edits}
\end{figure*}

Table~\ref{tab:detection} gives the development-cohort counts, and
Figures~\ref{fig:primary-detection} and~\ref{fig:primary-edits} show strength and strongest-window
length. Every marked sequence was found in both models and every condition. Each control produced at
most one positive sequence per model and condition, and the GENERator wrong-key positive is absent
after a deletion. With 192 prompts, the exact control intervals reach 2.868\% for one positive
prompt and 1.903\% for none. This cohort therefore cannot show whether the false-positive rate lies
below 1\%; the detection cohort addresses that question.

\begin{table*}[tp]
\centering
\caption{Positive detections in the development cohort, for each model across 192 held-out prompts
with two draws each. Correct-key and wrong-key columns refer to marked sequences. Each edit changes
exactly one base. The final two rows summarize unedited sequences at the prompt level to account for
paired draws; intervals are exact and two-sided. All marked prompts also had both draws detected in
every edit condition.}
\label{tab:detection}
\small
\setlength{\tabcolsep}{4pt}
\renewcommand{\arraystretch}{1.08}
\begin{tabular}{@{}lccc@{\hspace{12pt}}ccc@{}}
\toprule
& \multicolumn{3}{c}{\textbf{Carbon-500M}} & \multicolumn{3}{c}{\textbf{GENERator-v2 1.2B}} \\
\cmidrule(lr){2-4}\cmidrule(l){5-7}
Edit applied to the sequence & Correct key & Ordinary & Wrong key & Correct key & Ordinary & Wrong key \\
\midrule
None                 & 384/384 & 1/384 & 0/384 & 384/384 & 0/384 & 1/384 \\
One substitution     & 384/384 & 1/384 & 0/384 & 384/384 & 0/384 & 1/384 \\
One insertion        & 384/384 & 1/384 & 0/384 & 384/384 & 0/384 & 1/384 \\
One deletion         & 384/384 & 1/384 & 0/384 & 384/384 & 0/384 & 0/384 \\
\midrule
Prompts with a positive draw & 192/192 & 1/192 & 0/192 & 192/192 & 0/192 & 1/192 \\
Exact 95\% interval  & 98.10--100\% & 0.013--2.868\% & 0--1.903\% & 98.10--100\% & 0--1.903\% & 0.013--2.868\% \\
\bottomrule
\end{tabular}
\end{table*}

\subsection{Strength and the weakest sequence}
\label{supp:weakest}

\begin{figure*}[tp]
\centering
\makebox[\textwidth][l]{\small\textbf{Carbon-500M}}\par\vspace{0.15em}
\includegraphics[width=\textwidth]{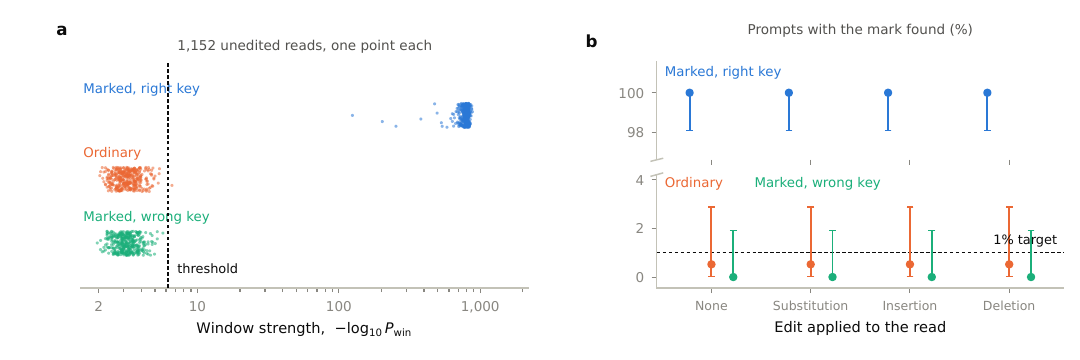}
\vspace{0.35em}
\makebox[\textwidth][l]{\small\textbf{GENERator-v2 1.2B}}\par\vspace{0.15em}
\includegraphics[width=\textwidth]{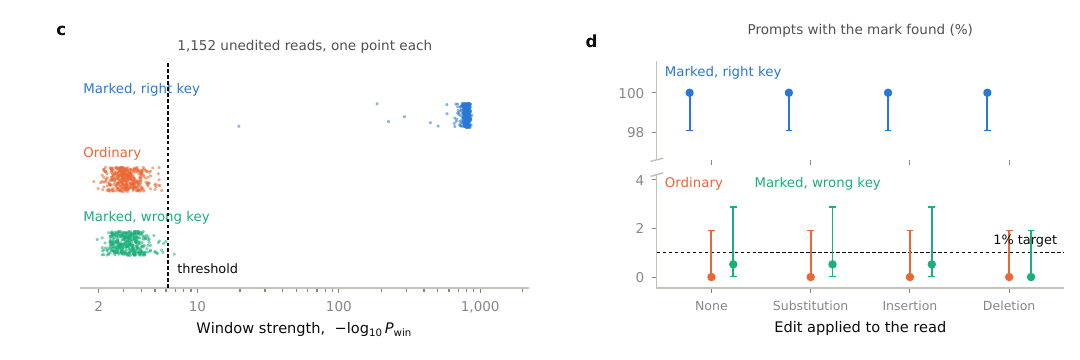}
\caption{\textbf{Position-independent detection in the development cohort.} \textbf{a,c}, Strength
of the best window in each unedited sequence, one point per sequence, on a log scale; in the panel labels,
``Marked, right key'' denotes marked sequences scored with the correct key, and ``reads'' denotes
test sequences. The dashed line is the threshold corrected for
the search over 16{,}136 windows. All marked sequences exceed the threshold. \textbf{b,d}, Proportion
of the 192 held-out prompts with positive detections, with exact two-sided 95\% intervals, on a
broken axis. All marked prompts were detected in all four edit conditions. The GENERator wrong-key
positive is absent after a deletion. With 192 prompts, the control intervals extend above the 1\%
target; Table~\ref{tab:second} and Figure~\ref{fig:detection} report the detection cohort. Panels
\textbf{a,b} show Carbon and panels \textbf{c,d} show GENERator.}
\label{fig:primary-detection}
\end{figure*}

Typical strength is nearly the same in both models. In the development cohort, the median unedited
marked sequence had strength 790.6 in Carbon and 798.8 in GENERator. Control strengths were
concentrated near or below the threshold, and the few positive controls lay only slightly above it
(Figure~\ref{fig:primary-detection}a,c). The weakest marked sequence differs more between the models.
In the development cohort, it had strength 124.8 in Carbon and 19.6 in GENERator. In the detection
cohort, the weakest sequences were 237.6 in Carbon and 123.8 in GENERator, and the medians were again close
(Table~\ref{tab:window}). The difference between weakest sequences therefore reflects single sequences, not
the models' typical margin.

The weakest detected GENERator sequence in the development cohort contained a highly repetitive
continuation. Of its
four-token contexts, 174 repeated and were excluded from scoring, which left 334 of a possible 508
tokens. Only 54.6\% of its mark bits were 1, against a median of 74\% in both models. Restoring all
508 scored tokens at that fraction would close only 9\% of the gap to Carbon's weakest sequence, so most
of the shortfall comes from weak signal in the tokens that remained. The tournament can raise the
fraction of 1s only where the model spreads probability across several tokens, which is consistent
with a low-entropy, repetitive stretch carrying little signal. We did not measure per-token entropy
to confirm this. We therefore find no evidence that typical detection strength differs between the
models.

\subsection{Aligned-detector diagnostic in GENERator}
\label{supp:aligned}

For GENERator, a separate diagnostic checked the analytic null with a detector given the known
token phase. Across the four evaluated lengths, 2, 6, 6, and 4 ordinary prompts were positive,
against about five expected at each length ($P = 0.25$--$0.87$). This result is compatible with
the analytic null, but it does not replace the full-search correction used for the
position-independent decision.

\subsection{Strength and window length by edit}
\label{supp:window}

Table~\ref{tab:window} gives the median strength of marked sequences in the detection cohort and, for
both cohorts, how often the full 3{,}072-base window carried the strongest evidence.

\begin{table*}[tp]
\centering
\caption{Strength and strongest-window length of marked sequences, by edit. Median strength is for the
detection cohort, over 3{,}088 sequences per model and condition. The full-window columns count sequences
whose strongest window was the full 3{,}072-base window, out of 384 per model and condition in the
development cohort and 3{,}088 in the detection cohort. Each edit changes exactly one base.}
\label{tab:window}
\small
\setlength{\tabcolsep}{4pt}
\renewcommand{\arraystretch}{1.08}
\begin{tabular}{@{}lcc@{\hspace{12pt}}cc@{\hspace{12pt}}cc@{}}
\toprule
& \multicolumn{2}{c}{\textbf{Median strength}} & \multicolumn{2}{c}{\textbf{Full window, development}} & \multicolumn{2}{c}{\textbf{Full window, detection}} \\
\cmidrule(lr){2-3}\cmidrule(lr){4-5}\cmidrule(l){6-7}
Edit applied to the sequence & Carbon & GENERator & Carbon & GENERator & Carbon & GENERator \\
\midrule
None             & 792.9 & 799.7 & 383 & 381 & 3{,}087 & 3{,}085 \\
One substitution & 777.6 & 783.8 & 383 & 381 & 3{,}087 & 3{,}085 \\
One insertion    & 441.8 & 443.1 & 203 & 200 & 1{,}724 & 1{,}732 \\
One deletion     & 437.7 & 438.5 & 232 & 227 & 1{,}663 & 1{,}671 \\
\bottomrule
\end{tabular}
\end{table*}

\end{document}